\documentclass[onecolumn,12pt,trackchanges]{aastex62mtl}
\usepackage[utf8]{inputenc}
\usepackage{natbib}
\usepackage{xcolor}
\usepackage{times}
\usepackage{enumitem}
\usepackage[bookmarks=false]{hyperref}

\usepackage{graphicx}
\bibpunct{[}{]}{,}{n}{}{;}

\usepackage{graphicx}
\usepackage{amssymb}
\usepackage{lineno}
\usepackage{changepage}

\hypersetup{
citecolor = {blue}
}

\begin{document}

%\begin{adjustwidth}{-2cm}{-2cm}

\title{{\huge Astro2020 State of the Profession White Paper} \\ ~ \\ \vspace{0.25in}
%\hspace{1cm}\title{{\huge Astro2020 State of the Profession White Paper}\hspace{1cm} \\ ~ \\ \vspace{0.25in}

{
%\begin{changemargin}{0cm}{0cm} %Apparently 0cm works...
\Large\bf NANOGrav Education and Outreach: Growing a Diverse and Inclusive Collaboration for Low-Frequency Gravitational Wave Astronomy
%\end{changemargin}
} \\ ~ \\
\vspace{12pt}
\hspace{-0.6in}\includegraphics[width=3in]{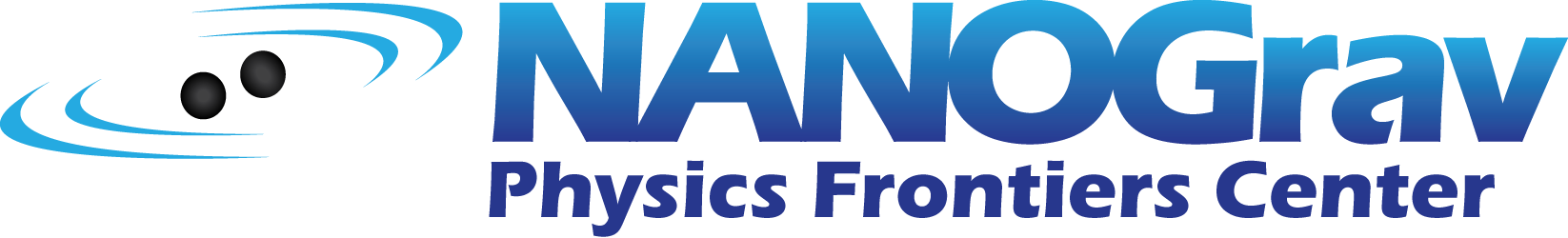}
%\hspace{-0.6in}\includegraphics[width=3in]{logo.png}
~ \\ \vspace{12pt}
{\large The North American Nanohertz Observatory for Gravitational Waves}
~ \\ ~ \\
{\rm  \today}
}

\shorttitle{NANOGrav Education and Outreach}
\shortauthors{NANOGrav Collaboration}

\section*{~~}

\vspace{-40pt}
%\vspace{-24pt}

%\end{adjustwidth}

{\bf Thematic areas:} Multi-messenger astronomy and astrophysics; Cosmology and fundamental physics; Formation and evolution of compact objects.

{\bf Contact author:} Timothy Dolch (NANOGrav Education and Public Outreach Chair), Timothy.Dolch@NANOGrav.org

%%%\vspace {-0.35in}

%%%\vspace {0.2in}

    {\bf Authors:}
P.~T.~Baker (WVU, Widener), H.~Blumer (WVU), A.~Brazier (Cornell), S.~Chatterjee (Cornell), B.~Christy (NDMU), F.~Crawford (F\&M), M.~E.~DeCesar (Lafayette), T.~Dolch (Hillsdale), N.~E.~Garver-Daniels (WVU), J.~S.~Hazboun (UW Bothell), K.~Holley-Bockelmann (Vanderbilt), D.~L.~Kaplan (UWM), J~.S.~Key (UW Bothell), T.~C.~Klein (UWM), M.~T.~Lam (WVU, RIT), N.~Lewandowska (WVU), D.~R.~Lorimer (WVU), R.~S.~Lynch (GBO), M.~A.~McLaughlin (WVU), N.~McMann (Vanderbilt), J.~Page (UAH), N.~Palliyaguru (AO), J.~D.~Romano (TTU), X.~Siemens (OSU, UWM), J.~K.~Swiggum (UWM), S.~R.~Taylor (Caltech/JPL, Vanderbilt), and K.~Williamson (WVU) for the NANOGrav Collaboration ($\sim 50$ institutions, 100$+$ individuals)
%\hspace{8cm}

%\hoffset=-0.8in
%\hoffset=0in

\begin{figure}%[h]
    
   %\centering
    \hspace{-0.55in}\includegraphics[width=7.6in]{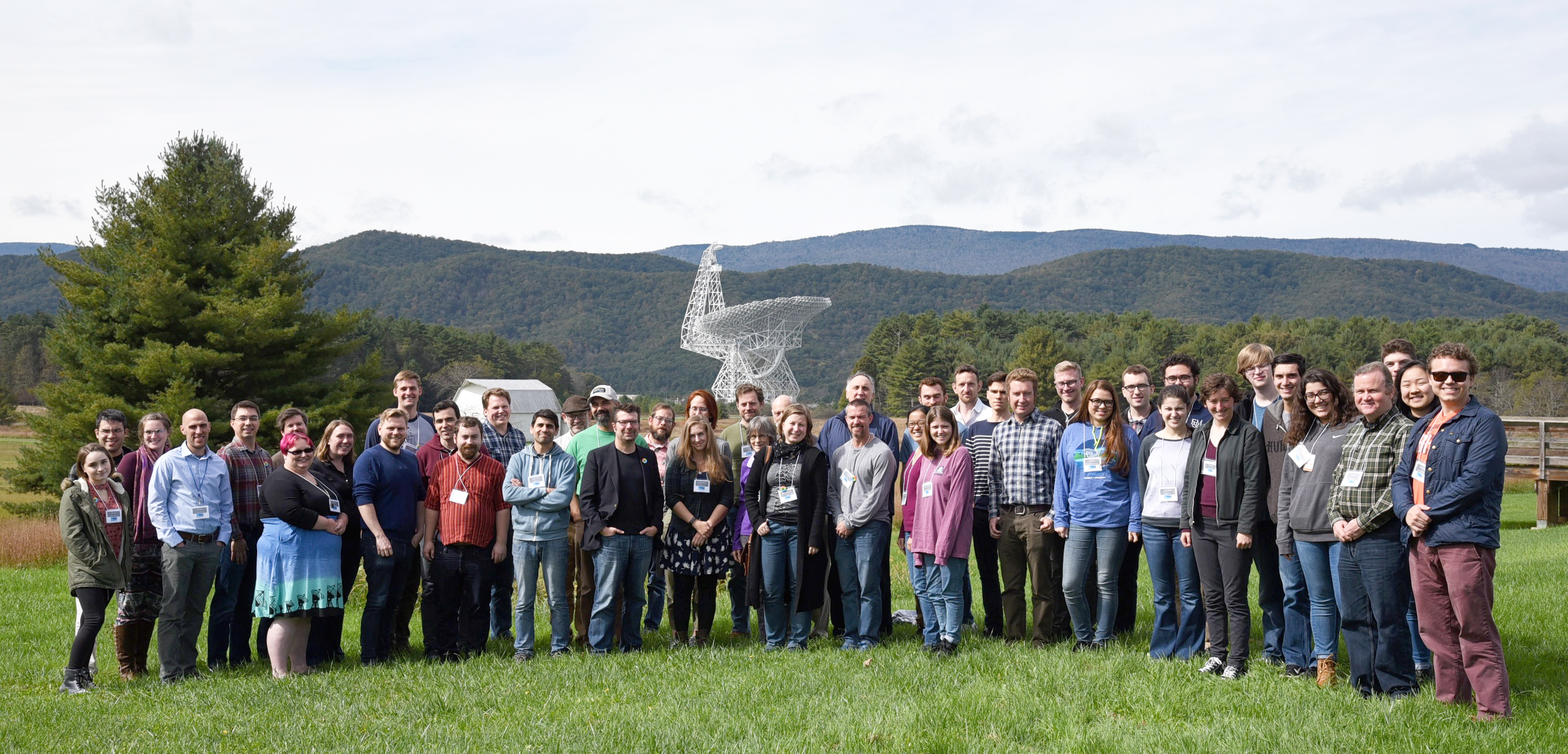}%[width=6.25in]{GBO_group.jpg}
    \label{fig:cover}
\end{figure}
%\hspace{4cm}

\newpage
\begin{center}
    {\bf Abstract}
\end{center}
\vspace{-12pt}
The new field of gravitational wave astrophysics requires a growing pool of students and researchers with unique, interdisciplinary skill sets. It also offers an opportunity to build a  diverse, inclusive astronomy community from the ground up. We describe the efforts used by the North American Nanohertz Observatory for Gravitational Waves (NANOGrav) NSF Physics Frontiers Center to foster such growth by involving students at all levels in low-frequency gravitational wave astrophysics with pulsar timing arrays (PTAs) and establishing collaboration policies that ensure broad participation by diverse groups. We describe and illustrate the impact of these techniques on our collaboration as a case study for other distributed collaborations. 

\section{Brief Introduction to NANOGrav}

\setcounter{page}{0} 
\pagenumbering{arabic}

The North American Nanohertz Observatory for Gravitational Waves (NANOGrav) collaboration is a National Science Foundation (NSF) Physics Frontiers Center dedicated to the detection of gravitational waves (GWs) at low ($10^{-9}-10^{-7}$ Hz)  GW frequencies and the characterization of the low-frequency GW universe. Our GW detectors are rapidly rotating neutron stars called  millisecond pulsars (MSPs) which act as incredibly precise cosmic clocks. Spread throughout the Galaxy, each pulsar--Earth ``arm'' of the detector is sensitive to perturbations to the space-time metric due to GWs with $\sim$month-to-decade-long periods. The most likely astrophysical sources in this regime are  supermassive binary black holes at the cores of galaxy mergers. The detection of low-frequency GWs will offer insights into the processes of galaxy evolution and structure formation not accessible by other means.

Simulations incorporating NANOGrav's sensitivity and the expected amplitude of the background due to supermassive binary black hole mergers indicate that the unique
correlated signature due to GWs  \citep{Hellings1983} should be detected in NANOGrav data within the next five years (see NANOGrav Science White Paper by Taylor et al. \citep{WP_Taylor}). In fact, the upper limits set on the stochastic background due to GWs in our most recent 11-yr dataset paper  are already astrophysically relevant, ruling out  parameter space for galactic bulge--black hole mass relationships \cite{nano11yr_gwb}.

The NANOGrav collaboration was formed in 2007 and received its first multi-institutional funding under from the NSF PIRE (Partnerships for International Research and Education) program (award number 0968296) in 2010. NANOGrav drafted by-laws and a constitution, ensuring its long-term presence as a coherent organization across many institutions in both the US and Canada (see \url{http://nanograv.org}). NANOGrav is also a member of the International Pulsar Timing Array (IPTA), along with partner collaborations in Europe (the European Pulsar Timing Array collaboration; EPTA) and Australia (the Parkes Pulsar Timing Array collaboration; PPTA).  IPTA datasets include  pulsars in both the Northern and Southern hemispheres, providing the wide range of angular separations ideal for GW detection \citep{IPTADR1}.

In 2015, NANOGrav was funded as an NSF Physics Frontiers Center  (award number 1430284), which facilitated dramatic growth in the number of participants and institutions. From 2007 to present, it has expanded from roughly a dozen members  to over 100 graduate students, postdocs, and senior personnel at roughly 50 institutions. We also involve  roughly 100 undergraduate students in NANOGrav work.    At the same time, the IPTA is currently expanding to  include collaborators in South Africa, India, and China.  Due to the need for a long-term investment in training NANOGrav members at all skill levels, as well as recognizing the need for public support of the Green Bank Observatory (GBO) and the Arecibo Observatory (our flagship facilities), education and public outreach have played an especially critical role in the history of our collaboration.

NANOGrav is a fertile environment for education and outreach due to  its deep integration of student training and research with its scientific goal of detecting GWs with pulsars. First, pulsar searching has long been uniquely suited to entry-level researchers because of the historical need for manual inspection of a copious number of candidate pulsar diagnostic plots, which are generated  at a few computational hubs and shared online with researchers at any institution. 
Second, many of our  institutions are undergraduate-only and several  have  large fractions of URM students. Third, 

the large observing programs for both pulsar searching and timing using Arecibo and the Green Bank Telescope (GBT) provide  tangible experiences   for students, enabling  intense pedagogical immersion outside the classroom. Fourth, the critical interaction of NANOGrav with  our IPTA partners means that we are an intrinsically diverse collaboration in which students can harness and continually improve important 21st-century skills in working in a distributed fashion across cultural barriers.

\begin{figure}[t]
    \includegraphics[width=\textwidth]{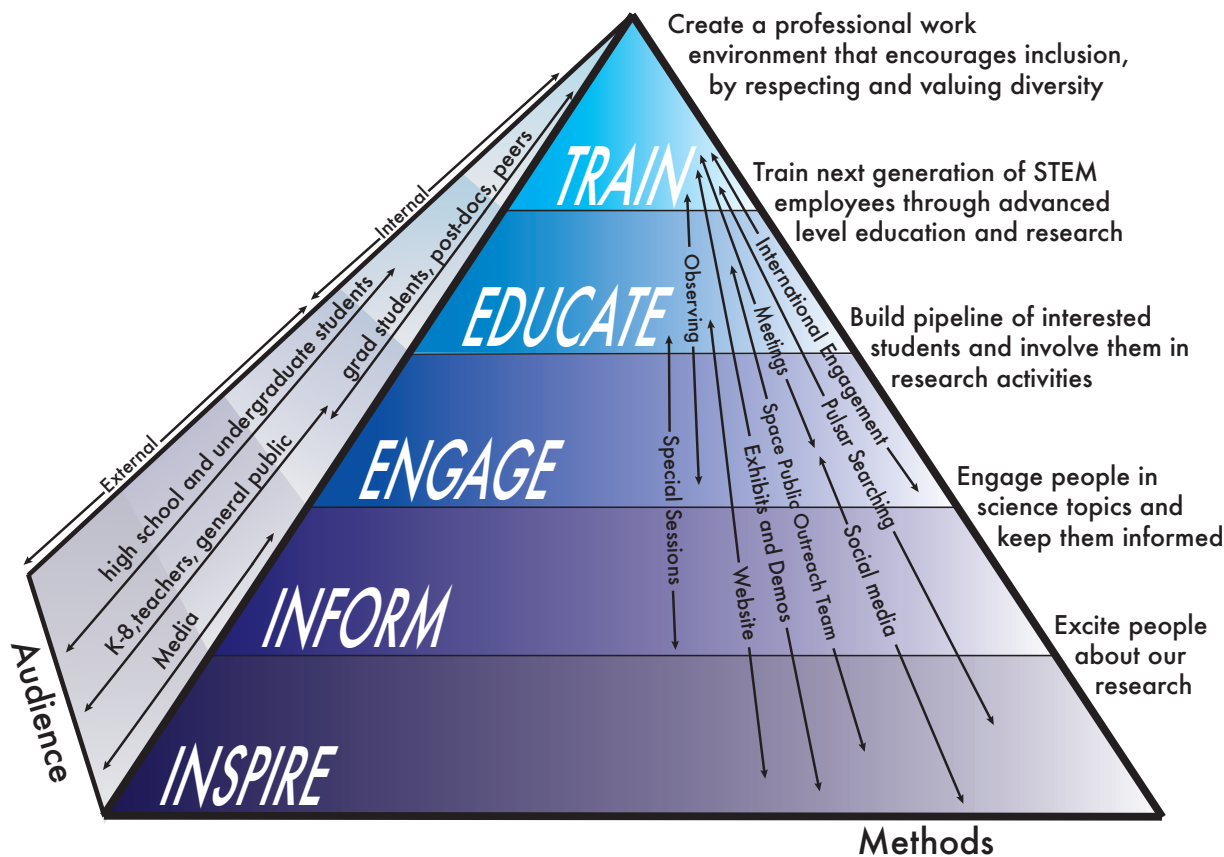}
    \caption{NANOGrav's EPO pyramid demonstrates the importance of initiating broad and inclusive outreach efforts if one wants to ultimately build a well-trained and diverse scientific community. Our efforts are designed to reach a variety of audiences at all levels through a range of initiatives. The sizes of the pyramid at the different levels reflect those of the intended audiences.   The dimensionality represents our multifaceted approach; while some initiatives target audiences solely outside or inside the collaboration, several are specifically designed to bridge the gap in a concerted effort to bring in new members at the ``Engage" level.
    }
    
       \label{fig:pyramid}
\end{figure}

\section{The Importance of Diversity in Science}
Aside from fundamental aspects of fairness and equity, numerous studies have shown that increasing diversity results in improved scientific output \citep{Chubin:2008,Page:2007}. However, while the past several decades have seen improvements, STEM fields generally and physics and astronomy in particular have significantly lagged in  representing the broader community \citep{Porter:2019aip,NCSES18}.  Broadening  representation  therefore requires substantive changes to undergraduate and graduate education \cite{Rudolph19,Miller19}, as well as thoughtful efforts to engage younger students to develop an interest in the field. It also necessitates careful training in inclusive practices  for senior and junior scientists and mentoring  and professional development to keep early-career researchers engaged and progressing.

Multi-institutional, distributed collaborations such as
NANOGrav are becoming increasingly more common and present unique opportunities and challenges. While such collaborations can (and should!) have policies that foster diversity, the demographics are somewhat limited by that of its member institutions, which may have  disparate   admissions and hiring processes. Furthermore, while anti-harassment policies play an important role in such a collaboration, enforcement of these policies must often happen at the individual institution level.  However, this diversity in experience can strengthen collaborations such as NANOGrav by allowing us to exchange and transfer knowledge and best practices widely, benefiting from work at a variety of institutions and scales.  Below we describe our efforts to engage the general public, educate students, and improve the experience within NANOGrav for all members that we believe are broadly applicable to similar collaborations.

\section{NANOGrav Outreach and Education Programs}\label{sec:programs}

To frame NANOGrav's efforts in the field of education and public outreach (EPO), we have devised the ``EPO pyramid" (Fig.~\ref{fig:pyramid}).    The left side of the pyramid shows our audiences, ranging from the general public/media,  to students at a variety of levels,  postdocs, and ultimately our peers. We also note whether these audiences are external  or internal to NANOGrav.  The  base in the front of the pyramid lists various methods, such as social media, public exhibits, Space Public Outreach Team (SPOT; \S~\ref{sec:spot}) presentations,
pulsar searching  by high-school and undergraduate students (\S~\ref{sec:stars}), and special sessions at scientific meetings.  Finally,  the vertical dimension  describes the   goals of  NANOGrav PFC EPO, starting with inspiring a broad range of people and gradually focusing effort to training students and junior researchers for direct involvement in NANOGrav.  Throughout the pyramid various specific efforts are shown, along with the intended audience and level of engagement.

Below we give specific descriptions of representative NANOGrav EPO programs at various levels of the EPO pyramid.

\subsection{INSPIRE: Space Public Outreach Team (SPOT)}
\label{sec:spot}
The SPOT program provides a framework for training undergraduate students to present science talks and hands-on demonstrations to a variety of K--12 audiences \cite{spot}. The program was initiated at Montana State University in 1996. In 2014, a second SPOT program was initiated in West Virginia and in 2016 a NANOGrav-wide SPOT program was established. It has since become an integral part of NANOGrav PFC outreach efforts and also provides valuable professional development for students.

\begin{figure}
\centering
\includegraphics[width=\textwidth]{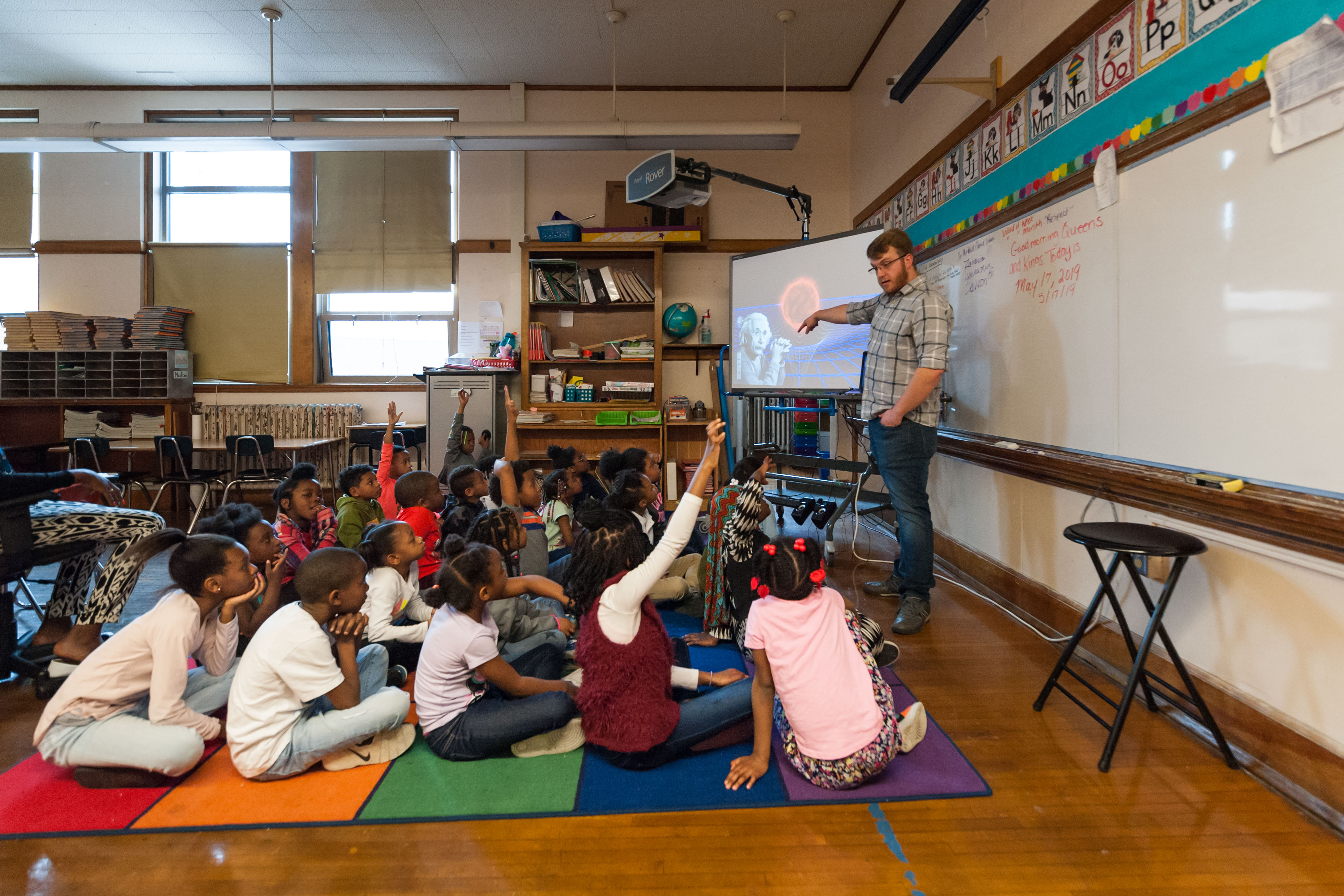} %0.3
\caption{\footnotesize
Graduate student Alex McEwen from University of Wisconsin-Milwaukee (and former NANOGrav undergraduate researcher at WVU) gives a SPOT presentation to second grade students at Hartford Avenue University School in Milwaukee, Wisconsin.}

\end{figure}

SPOT members give the presentation ``Tuning In to Einstein's Universe,'' an interactive 45-minute slideshow presentation that is targeted to a middle-school age audience but is adaptable to audiences of all ages.\footnote{See an example presentation on YouTube at \url{https://youtu.be/AmJ7eDouQFU} and download a tarred collection of presentation materials from \url{http://docdb.nanograv.org/?r=466&k=1fd70983f9}.} The presentation builds on the history of astronomy from Galileo's time to show how gravity can open a  new window to our understanding of the universe, addressing questions such as: How are gravitational waves made? What are pulsars? How do we use pulsars to detect them? Where is this research happening in your home state and how can you become involved? Different versions of the presentation highlight the unique contributions of the individual states where NANOGrav research takes place. To date, 113 SPOT presentations have been given by 42 NANOGrav members to a total of 5,611 audience members. Evaluation of the SPOT program shows that presenters learn valuable skills in science communication and teamwork while audiences demonstrate a greater understanding of GW astronomy and increased interest in exploring science careers.

\subsection{INFORM: General Public}

NANOGrav has reached the general public through a variety of avenues. NANOGrav PFC members have given over 100 public talks over the past several years. In addition to our main website, we have created the outreach website \url{http://explore.nanograv.org}. Through many public talks and several popular articles, we have reached wide audiences, ranging from middle- and high-school students \citep{frontiersin} all the way to amateur astronomers\footnote{\url{http://nanograv.org/assets/files/ab2018-161\%20May2018-Dolch.pdf}}. The University of Wisconsin-Milwaukee has also developed a number of exhibits that we integrated into a ``MakerFaire" display, which has been shown at the  MakerFaire Milwaukee since 2016.  The exhibits, engaging students in their design and construction, draw heavily from modern techniques like 3D printing and Arduino microcontrollers. Many of the exhibit materials are now regularly used at NANOGrav outreach events around the country. NANOGrav members also participated in a     ``STEM and the Arts Film Festival'' in West Virginia in March/April 2018 and presented demos involving NANOGrav science to over 800 middle-school students. These students viewed the ``little green men'' documentary\footnote{\url{http://www.lgmfilm.com/}} (funded through NSF AISL award number 1137082) about the Pulsar Search Collaboratory  (\S~\ref{sec:psc}).

We are actively engaged in social media on Twitter (1202 followers), Facebook (1494 followers), Instagram (450 followers), a YouTube channel (239 subscribers), and have implemented consistent branding across social media. We feel that using social media can help us reach a more diverse audience in terms of gender, race, age, and geographic location. For example, Facebook boasts $\sim2.3$ billion monthly active users. Both Instagram and Twitter have younger user bases, primarily in the 13-17 and 18-29 age ranges, and Instagram reports a higher fraction of female versus male users. 

In addition, online interactions can sometimes lead to real-world connections. Recently, we advertised our student workshop held at UW Bothell on Facebook and Twitter. One female student who was not enrolled in a NANOGrav institution saw the post, signed up for the student workshop, and attended it in Spring 2019. We consider this a step forward in terms of recruiting a more diverse student population for future workshops.  

\newpage

\subsection{ENGAGE: Pulsar Search Collaboratory (PSC)}
\label{sec:psc}

\begin{figure}[t]
\centering
\includegraphics[width=0.49\textwidth]{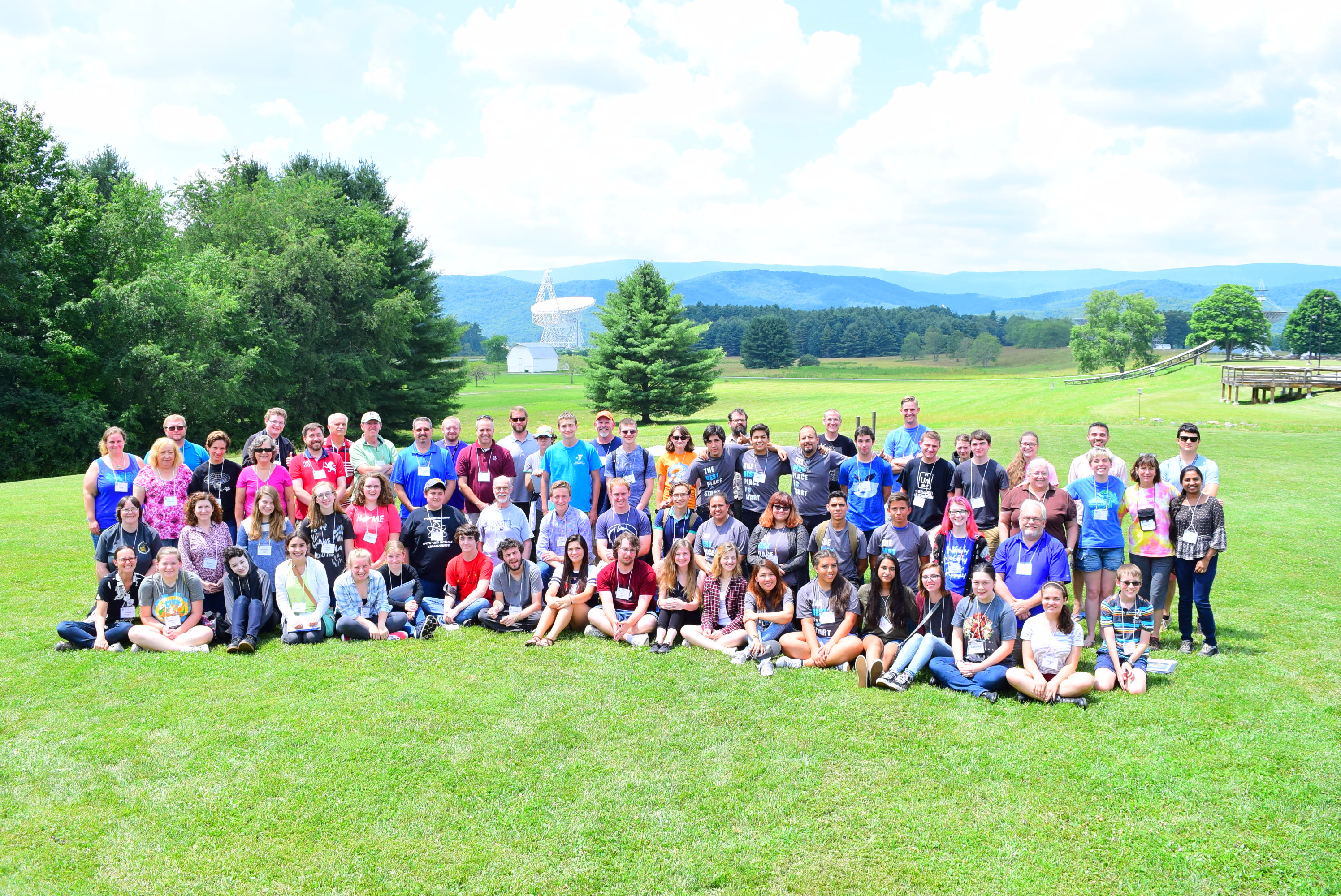} %0.3
\includegraphics[width=0.49\textwidth]{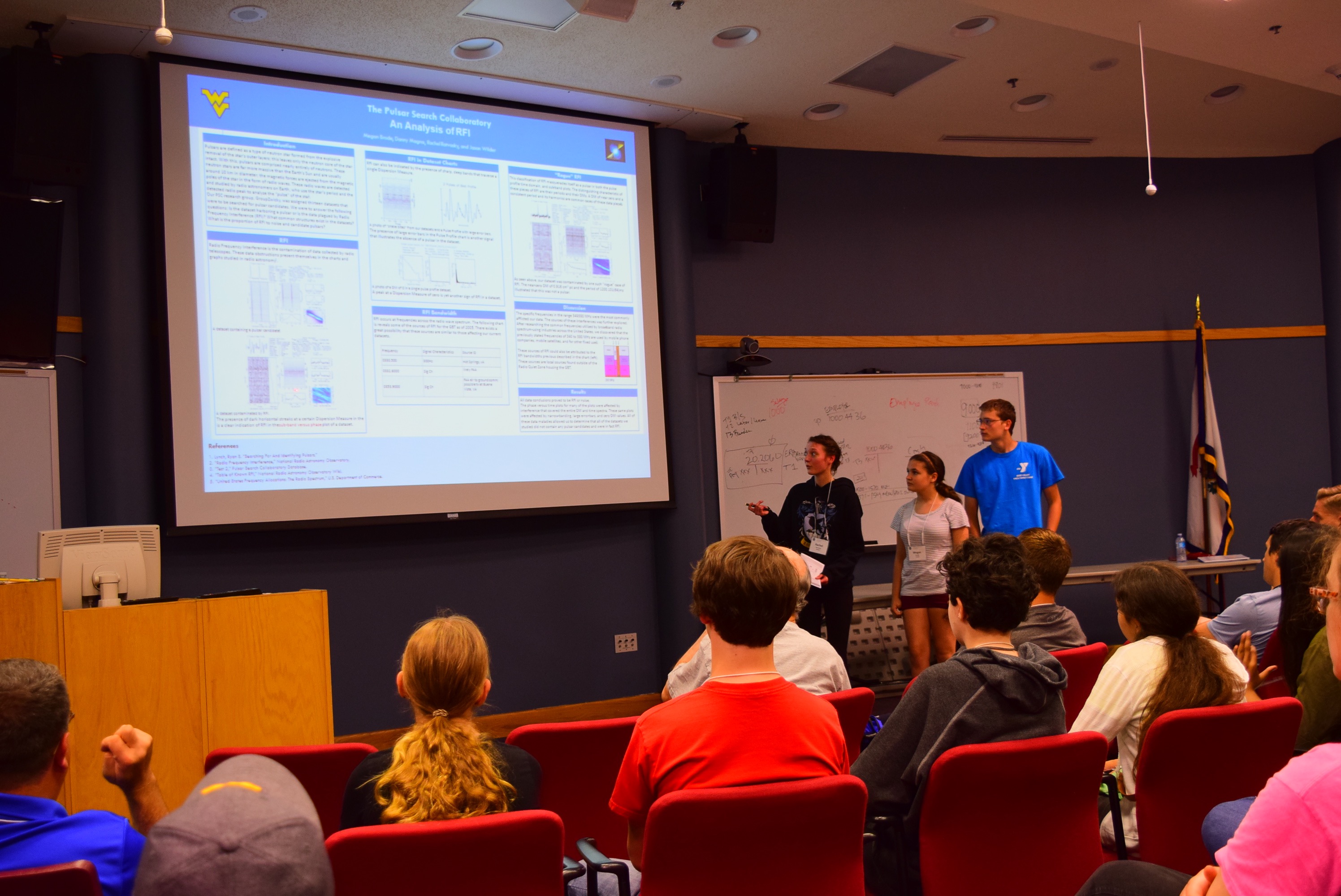} %0.1
\caption{\footnotesize
PSC participants at the camp in Green Bank (left) and students presenting their work during the camp (right).}
%\vspace{-5mm}
\end{figure}

The PSC program, originally funded through the NSF AISL program (award number 0737641), began in 2007 as a collaboration between GBO and West Virginia University (WVU). The PSC engages middle and high-school students and their teachers in  searching for pulsars in data collected with the GBT and set aside solely for student analysis.  The goal of the PSC program is to promote STEM interests and careers and encourage student use of information technologies through immersing the students in an authentic, positive learning environment designed to build a sense of belonging and competency \cite{rosenpsc}. A recent expansion of the program, funded through NSF award numbers 1516512 and
1516269 has fostered the  development of online workshops and the involvement of undergraduate mentors at  hub institutions across the country, providing a valuable professional development opportunity for undergraduates \cite{williamsonpsc}.

 PSC participants learn about radio astronomy, pulsars, radio frequency interference, pulsar timing, gravitational waves, programming, and big data analysis. After the training, the participants complete online certification tests to gain access to the GBT data. At the end of every academic year, the most active participants  qualify to attend Capstone events held at different hub institutions. At these several-day events, they present posters on their research, listen to talks by various researchers, and tour campus facilities and laboratories. Students and teachers also attend a one-week  camp in Green Bank. Thus far, 2,432 students, 106 teachers, and 80 undergraduate students in 18 states and from 195 schools have been exposed to research through the PSC. Of those students, 191 have traveled to the GBO to participate in the summer camps. 

The participation is diverse, with roughly 50\% from rural counties and roughly  50\% female. The program had a positive impact on the middle and high-school students according to multiple measures, in particular, on their understanding of the nature of scientific inquiry and motivation to pursue STEM career paths. In addition, they showed higher levels of science identity, self-efficacy, and STEM intentions compared to a control group of students who did not participate in the PSC. Many of the former PSC students are now undergraduate students at NANOGrav institutions and  participate in the PSC as mentors. The high-school teachers also showed significant increases in their competency and confidence in teaching science and the research process. They also indicated that they planned to implement PSC research in their classroom and incorporate PSC research activities and research knowledge into curriculum discussion. 

Furthermore, undergraduate students who served as PSC mentors showed  statistically significant increases in their research knowledge and confidence, and in their research, leadership, communication, and collaboration skills. Additionally,  93\% of the undergraduate mentors were majoring in STEM fields  shared that participation encouraged them to continue majoring in STEM and pursue STEM careers.

The PSC has also been a great scientific success. Seven pulsars, including one MSP  and one double neutron star binary, have been discovered by PSC students so far. Future discoveries may be used for fundamental advances such as for testing of general relativity, constraining neutron star masses, and detecting gravitational waves. In this way, the PSC students are not ``playing the part'' of scientists but truly contributing to the work of the collaboration and, more broadly, the scientific community.

\subsection{EDUCATE: Student Teams of Astrophysics ResearcherS (STARS)}
\label{sec:stars}
The NANOGrav STARS program engages undergraduate students at NANOGrav institutions and serves as a point of entry for students interested in pulsar and GW research  at various stages of their undergraduate careers. This program has two concurrent goals. The first is educating students in the basics of GW science and developing teamwork, leadership, and speaking/presenting skills (``soft skills''). 
The second goal is to provide meaningful student research experiences, to network with more senior researchers and mentors, and to make tangible contributions to the core science of NANOGrav (e.g., pulsar discoveries, observing) in the form of team-based projects as well as individualized research projects under the supervision of senior mentors. This is enhanced by face-to-face exposure to senior personnel and postdocs on weekly telecons.

The STARS program involves weekly videoconferences during which students from different institutions present short talks, usually on their research work but sometimes on recent astronomy discoveries. This fosters the development of presentation skills and also provides valuable information for junior students who may be interested in getting involved in research at a later stage. Faculty and postdocs also attend these telecons and present talks both on research and on topics such as paper writing, graduate school applications, and proposal preparation.

\begin{figure}[t]
\centering
\includegraphics[scale=0.25]{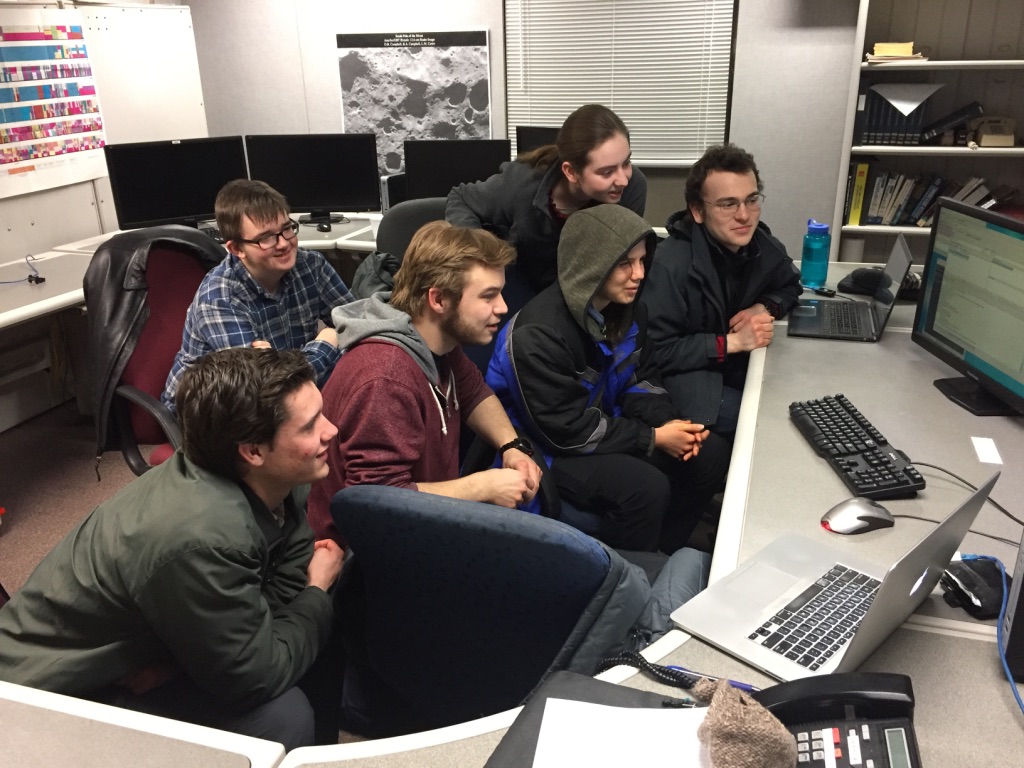} %0.3
\includegraphics[scale=0.08]{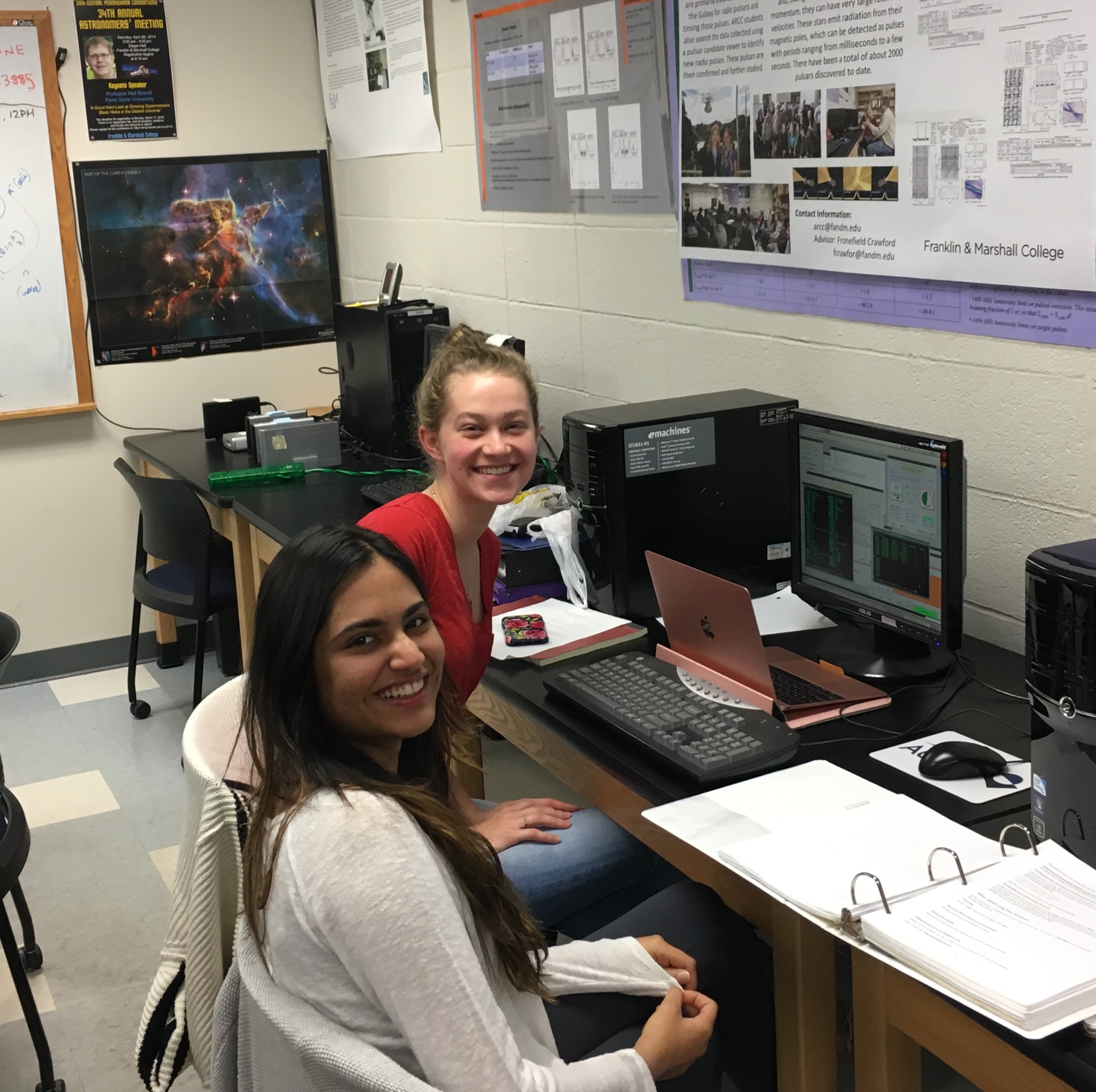} %0.1
\caption{\footnotesize
STARS students observe at the Green Bank Telescope control room (left) and remotely for Arecibo timing observations (right).           
\label{fig:observers}}

\end{figure}

The STARS teams at each NANOGrav institution also conduct critical science for the collaboration. Undergraduate observers use Arecibo and the GBT either remotely or on site to conduct a sizable fraction of our routine timing observations (see Fig.~\ref{fig:observers}). 
The observing contribution of the undergraduate students is critical  given the $\sim$1200\,hr/year of observing time allotted to Arecibo and the GBT for NANOGrav observations (i.e.  $\sim$twice-monthly observations of 76 MSPs). In one recent semester, STARS undergraduates conducted 40\% of all Arecibo timing observations for NANOGrav. In addition to long-term timing, STARS groups also become involved with pulsar searching with the PALFA \citep{palfa}, AO327 \citep{deneva}, and GBNCC \citep{stovall} pulsar surveys. The pulsar discoveries they make are essential for GW detection, as the signal-to-noise  of the GW background  amplitude scales linearly with the number of pulsars 
regularly timed in the PTA \citep{siemens}. STARS students use the CyberSKA\footnote{\url{https://ca.cyberska.org/}} online software hosted by McGill University to visually inspect candidates and some STARS groups conduct pulsar observations to follow-up on good candidates. Together, these experiences are both essential to NANOGrav science and provide a rich pedagogical experience for students. 

In 2019, NANOGrav had 90 undergraduate students involved in NANOGrav research, including 60 regularly attending STARS telecons, across  eight institutions. {77 students have participated in research abroad (see the IRES program below) and {11} NANOGrav publications in the past two years included undergraduate co-authors.}

We note that a significant fraction of NANOGrav STARS students come from underrepresented groups (roughly 45\% female and 15\% URM). Therefore,   we believe that this  program can serve as a critical pipeline for recruiting and retaining students from all backgrounds who are interested in participating in this work. 

Evaluation of our undergraduate research program shows that undergraduate students who conducted research showed  statistically significant growth in their understanding of astrophysics and pulsars and in their skills related to astrophysics data analysis. They also showed a statistically significant increase in their confidence in understanding scientific concepts, designing research projects, and communicating research results.
Interestingly,
these effects were amplified for URM and female students. URM respondents indicated higher levels of understanding than non-URM respondents and
both URM and female respondents indicated higher levels of confidence than non-URM and male respondents.

In addition, undergraduate students found both research and career mentorship to be very useful, with URM respondents perceiving the mentorship they received as more useful than non-URM respondents. In written feedback, respondents also indicated they received encouragement and gained career knowledge and skills by working with their mentors.

Furthermore, we found that 72\% of undergraduate students who have participated in the STARS program are highly or very highly interested in pursuing a career in physics/astrophysics and 69\% were highly or very highly interested in pursuing a graduate degree in physics/astrophysics. This is born out by longitudinal studies which show that roughly 80\% of former STARS students remain engaged in STEM fields after participation.

\begin{figure}[t]
\centering
\includegraphics[width=\textwidth]{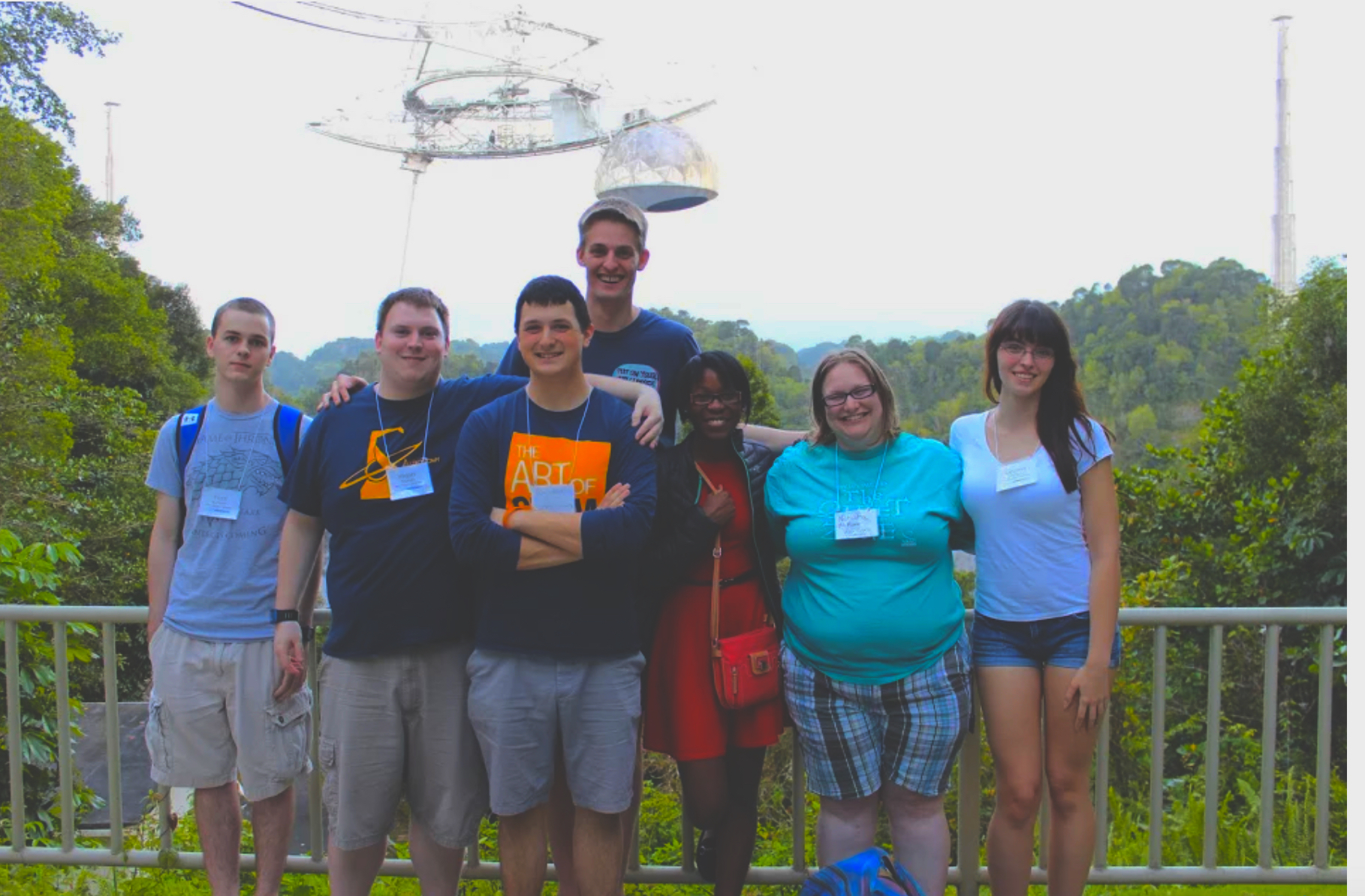} %0.3
\caption{\footnotesize
NANOGrav undergraduates at a student workshop at Arecibo Observatory in Puerto Rico.}
%\vspace{-5mm}
\end{figure}

\newpage

\subsection{EDUCATE: Undergraduate Education}
\label{sec:ugrad}

Several NANOGrav online resources have been developed in order to educate undergraduates in PTA science, including \texttt{\url{http://gravcalc.org}} and \texttt{\url{http://simulator.nanograv.org}}. These are helpful for students who are not already part of the training process within a PSC institution. We have also made online resources for tabletop demos available, including a paper on  a metronome-based pulsar timing demo, published in the American Journal of Physics  \cite{2018AmJPh..86..755L} and also published in the Mexican popular science magazine ``Conversus'' in 2016 \cite{conversus}.

Additionally, the NANOGrav EPO working group coordinates student workshops at NANOGrav and IPTA meetings and creates Jupyter notebooks available publicly online. We have also developed material for a semester-long course in pulsar timing, given at UWM in Fall 2018. The class began with an introduction to programming (conducted through an online partner organization) alongside classroom lectures on astronomy.  These were paired with active-learning computer lab sessions led by senior undergraduates.  By the end of the semester the students had all successfully determined timing solutions for newly-discovered pulsars, with the results published with undergraduate first-authorship \cite{Aloisi19}.  The course materials, including online computer lab exercises, are available for the NANOGrav collaboration and the broader IPTA collaboration. {A similar course was also taught at WVU.}

\begin{figure}[t]
\centering
\includegraphics[width=0.49\textwidth]{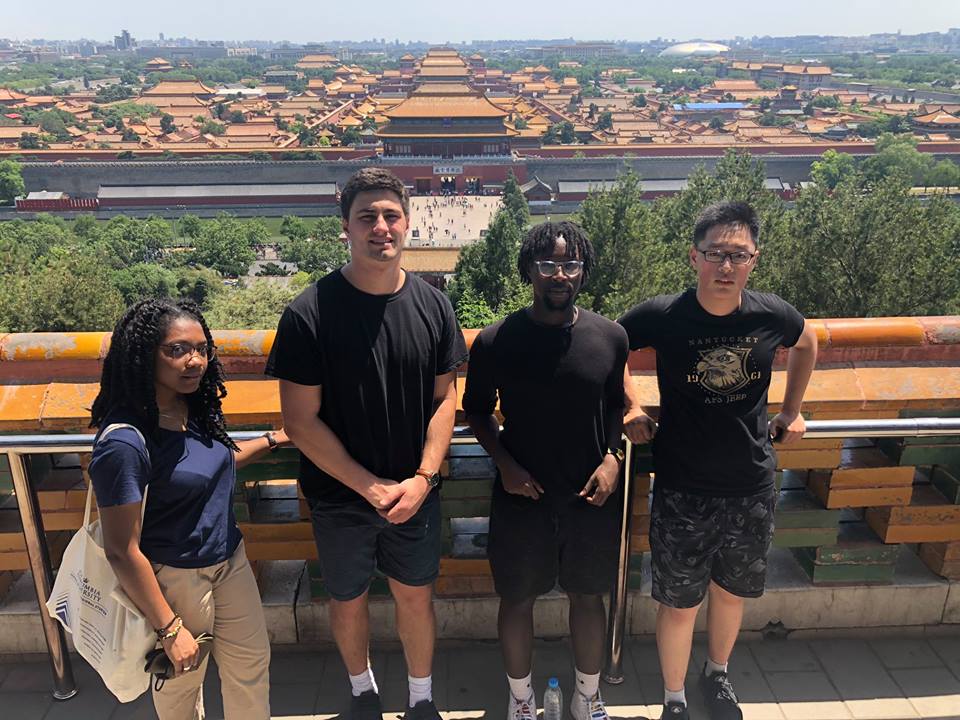} %0.3
\includegraphics[width=0.49\textwidth]{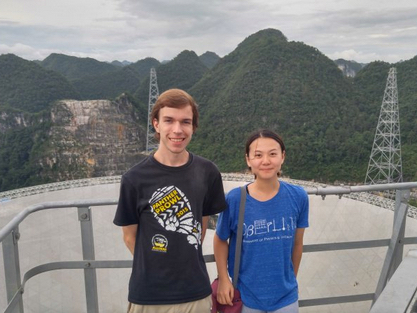} %0.1
\caption{\footnotesize
IRES undergraduate students in Beijing (left) and at the Five-hundred-meter Aperture Spherical radio Telescope (FAST) in China (right).}
%\vspace{-5mm}
\end{figure}

\subsection{TRAIN: IRES Program}
\label{sec:ires}

The NANOGrav IRES (International Research Experiences for Students)  program (which evolved out of our earlier PIRE program), funded through NSF award number 1658632, engages undergraduate students at NANOGrav institutions in research abroad with mentors   conducting NANOGrav-related research. This REU-style internship typically lasts for 10 weeks in the summer at the foreign institution. From 2010--2016, 47 students participated in research abroad experiences through the PIRE program and from 2017--2019, 30 students have participated through the IRES program. Of these IRES students, 43\% were female and 26\% are from URM groups. In total, students have been hosted at  16 institutions in 11  countries. In addition to providing valuable student training in both NANOGrav science and international communication and collaboration, this program  strengthens the ties between NANOGrav and other IPTA groups, an important consideration as the search for GWs engages more groups and facilities worldwide.

The IRES experiences are advertised on the NANOGrav STARS telecons, with students who have participated in such experiences presenting talks on the telecons. This also amplifies the impact of the IRES funding by educating a larger group of students about the benefits of international collaboration. Students who have participated in the program demonstrate increased knowledge in multiple research areas and statistically significant increases in  skills and confidence in international collaboration. They also demonstrated an increased  desire to work with scientists from other countries. Finally, all participants showed high or very high interest in pursuing a graduate degree or career in a STEM field and shared that the research abroad experience helped prepare them for graduate school or a career in a STEM field.

%%%\newpage

\subsection{TRAIN: Postdoctoral Mentoring}

The increased growth in the new field of GW astrophysics requires a larger number of  faculty members and scientists trained in both the relevant science and communication and collaboration skills. To address this need, NANOGrav 
initiated a targeted postdoctoral mentoring program two years ago. Each postdoctoral NANOGrav member is assigned a mentoring team of three senior personnel (their local advisor and two other NANOGrav Senior Personnel). The postdocs meet with their mentoring team twice yearly to discuss career goals, progress towards those goals, and ways in which the collaboration can assist postdocs in meeting them. These sessions ensure that postdocs are receiving thoughtful mentoring that is specifically tailored to their career goals.

We have recently expanded the postdoc mentoring program with an annual Career Webinar series, including presentations from colleagues who have gone on to careers inside and outside of academia. Such a series was successfully implemented in 2018 and continues to be hosted annually for postdocs and graduate students.

\section{Collaboration-Level Support of Climate and Diversity}
\label{sec:climate}

As a relatively new collaboration at the forefront of a new science frontier, we are aware of our responsibility to foster a diverse and inclusive culture. We have therefore developed policies to ensure that all members work in a supportive and equitable environment.  This begins with the onboarding of new members, where we attempt to instill the best of ``NANOGrav culture" and lead by example.  Over roughly a decade, we have developed an anti-harassment and diversity policy which all new members must read and abide by as part of their membership approval process. This is accompanied and reinforced by regular presentations on topics such as harassment, unconscious bias, microaggressions, and imposter syndrome at every one of our semi-annual collaboration meetings.  These presentations are recorded and membership renewal requires collaboration members to watch them offline if they are unable to attend the meeting in-person.

As a distributed collaboration, much of our communication is via telecons. We have found small initiatives such as
requiring members to speak only after raising their hand and being called on by the moderator to make a large differences towards inclusivity and ensuring that all viewpoints (not only the loudest!) are heard.
At our semi-annual collaboration meetings as well as any meeting that we co-sponsor (such as the annual IPTA meetings) we follow best practices to ensure the experience for all attendees is positive and welcoming.  This includes an explicit provision for junior members to ask questions before senior members and hosting student-only debriefing sessions. 

In addition, we have recently formed a Climate and Equity committee within NANOGrav to help monitor and assess structural issues within the collaboration that may contribute to an unhealthy  climate.  This committee does not investigate specific interpersonal disputes but is tasked with formulating policies and practices that ensure that all NANOGrav members have the resources and support they need to succeed.  Examples include  distributing best practices for writing/reading letters of recommendation, and adjusting meeting formats  and arranging for training  sessions at each meeting (as discussed above).  
We also connect to broader groups such as the Multimessenger Diversity Network\footnote{\url{https://icecube.wisc.edu/news/view/640}} and the APS Committee on the Status of Women in Physics\footnote{\url{https://www.aps.org/about/governance/committees/cswp/index.cfm}.} to learn from others and share best practices.

\section{Continuation and Integration of Outreach Programs over the Next Decade}\label{sec:integration}

We expect continued growth in the collaboration over the next decade. This  presents the challenge of preserving our inclusive culture by communicating expectations clearly and offering additional training opportunities targeted to diverse audiences. It also represents an incredibly valuable opportunity to involve even more participants in our science and to dramatically broaden our outreach efforts.

We plan to continue the initiatives described in this document and, in most cases, expand their demographic and geographic reach. As an example,
one of our short-term goals is to establish PSC and SPOT programs in Puerto Rico. The PSC program would involve middle and high-school students in analyzing Arecibo search data and the SPOT program would involve undergraduates from multiple institutions in giving talks across the island. These projects will naturally target minority-serving institutions and underrepresented minority (URM) groups in Puerto Rico. They also have the scientific benefit of exploiting the uniquely sensitive Arecibo surveys to increase our sensitivity to GWs through the discovery of additional MSPs.

\section{Summary}

No one document can provide a complete ``recipe'' for building a viable and dynamic education and outreach program. However, we hope that this description of NANOGrav initiatives over the past decade can serve as a case study of how such efforts can be incorporated into the structure of distributed and growing collaborations over the next decade. We urge other research communities to similarly document their  practices so that we can benefit from our shared experiences to mindfully build research communities that foster the involvement of a diverse range of participants. These actions will undoubtedly result in more effective science and optimize discovery potential over the next decade. In addition, broad outreach efforts  are also essential to ensure that scientific investments remain a priority for  US citizens, enabling cutting-edge science and preserving US competitiveness in the global research arena.

\section{Acknowledgments}

The NANOGrav project receives support from National Science Foundation (NSF) Physics Frontiers Center award number 1430284. The Arecibo Observatory is a facility of the National Science Foundation operated under cooperative agreement by the University of Central Florida in
alliance with Yang Enterprises, Inc. and Universidad Metropolitana.

The Green Bank Observatory is a facility of the National Science Foundation operated under cooperative agreement by Associated Universities, Inc. We thank SmartStart  and in particular Aubrey Roy for their evaluation of NANOGrav's program.

\newpage
%\singlespace
\bibliographystyle{unsrt}
\bibliography{references}

\begin{thebibliography}{10}

\bibitem{Hellings1983}
R.~W. {Hellings} and G.~S. {Downs}.
\newblock {Upper limits on the isotropic gravitational radiation background
  from pulsar timing analysis}.
\newblock {\em \apjl}, 265:L39--L42, February 1983.

\bibitem{WP_Taylor}
Stephen {Taylor}, Sarah {Burke-Spolaor}, Paul~T. {Baker}, Maria {Charisi},
  Kristina {Islo}, Luke~Z. {Kelley}, Dustin~R. {Madison}, Joseph {Simon}, Sarah
  {Vigeland}, and {Nanograv Collaboration}.
\newblock {Supermassive Black-hole Demographics \& Environments With Pulsar
  Timing Arrays}.
\newblock In {\em \baas}, volume~51, page 336, May 2019.

\bibitem{nano11yr_gwb}
Z.~{Arzoumanian}, P.~T. {Baker}, A.~{Brazier}, S.~{Burke-Spolaor}, S.~J.
  {Chamberlin}, S.~{Chatterjee}, B.~{Christy}, J.~M. {Cordes}, N.~J. {Cornish},
  F.~{Crawford}, et~al.
\newblock {The NANOGrav 11 Year Data Set: Pulsar-timing Constraints on the
  Stochastic Gravitational-wave Background}.
\newblock {\em \apj}, 859(1):47, May 2018.

\bibitem{IPTADR1}
J.~P.~W. {Verbiest}, L.~{Lentati}, G.~{Hobbs}, R.~{van Haasteren}, P.~B.
  {Demorest}, G.~H. {Janssen}, J.~B. {Wang}, G.~{Desvignes}, R.~N. {Caballero},
  and M.~J. {Keith}.
\newblock {The International Pulsar Timing Array: First data release}.
\newblock {\em \mnras}, 458(2):1267--1288, May 2016.

\bibitem{Chubin:2008}
D.~E. {Chubin} and S.~M. {Malcolm}.
\newblock {Making a Case for Diversity in STEM Fields}.
\newblock {\em Inside Higher Ed}, 2008.

\bibitem{Page:2007}
Scott {Page}.
\newblock {\em The Difference: How the Power of Diversity Creates Better
  Groups, Firms, Schools, and Societies}.
\newblock Princeton University Press, 2007.

\bibitem{Porter:2019aip}
Ann~Marie Porter and Rachel Ivie.
\newblock {Women in Physics and Astronomy, 2019}.
\newblock Technical report, AIP Statistical Research Center, College Park, MD,
  March 2019.

\bibitem{NCSES18}
{National Science Foundation, National Center for Science and Engineering
  Statistics, Directorate for Social, Behavioral and Economic Sciences}.
\newblock {Doctorate Recipients from U.S. Universities: 2017}.
\newblock {\em Special Report NSF 19-301}, Alexandria, VA. Available at
  \nsfurl, 2018.

\bibitem{Rudolph19}
Alexander {Rudolph}, Gibor {Basri}, Marcel {Ag{\"u}eros}, Ed~{Bertschinger},
  Kim {Coble}, Megan {Donahue}, Rachel~L. {Ivie}, Jackie {Monkiewicz},
  Christine {Pfund}, and Julie {Posselt}.
\newblock {Final Report of the 2018 AAS Task Force on Diversity and Inclusion
  in Astronomy Graduate Education}.
\newblock In {\em Bulletin of the American Astronomical Society}, volume~51,
  page 0101, Jan 2019.

\bibitem{Miller19}
Casey~W. Miller, Benjamin~M. Zwickl, Julie~R. Posselt, Rachel~T. Silvestrini,
  and Theodore Hodapp.
\newblock Typical physics ph.d. admissions criteria limit access to
  underrepresented groups but fail to predict doctoral completion.
\newblock {\em Science Advances}, 5(1), 2019.

\bibitem{spot}
K.~{Williamson}, A.~D. {Jardins}, I.~{Grimberg}, S.~L. {Larson}, J.~{Key},
  M.~B. {Larson}, S.~A. {Heatherly}, D.~{McKenzie}, and T.~B. {Littenberg}.
\newblock {The Space Public Outreach Team (SPOT): Adapting a successful
  outreach programme to a new region}.
\newblock {\em Communicating Astronomy with the Public Journal}, 16:8, Dec
  2014.

\bibitem{frontiersin}
Stephen~R. {Taylor}.
\newblock {Catching Gravitational Waves With A Galaxy-sized Net Of Pulsars}.
\newblock {\em arXiv e-prints}, page arXiv:1906.07568, Jun 2019.

\bibitem{rosenpsc}
R.~{Rosen}, S.~{Heatherly}, M.~A. {McLaughlin}, R.~{Lynch}, V.~I. {Kondratiev},
  J.~R. {Boyles}, M.~{Wilson}, D.~R. {Lorimer}, and S.~{Ransom}.
\newblock {The Pulsar Search Collaboratory}.
\newblock {\em Astronomy Education Review}, 9(1):010106, Jan 2010.

\bibitem{williamsonpsc}
Kathryn {Williamson}, Maura {McLaughlin}, John {Stewart}, Duncan {Lorimer},
  Harsha {Blumer}, Cabot {Zabriskie}, Sue~Ann {Heatherly}, and Ryan {Lynch}.
\newblock {The Pulsar Search Collaboratory: Expanding Nationwide}.
\newblock {\em The Physics Teacher}, 57(3):156--158, Mar 2019.

\bibitem{palfa}
J.~M. {Cordes}, P.~C.~C. {Freire}, D.~R. {Lorimer}, F.~{Camilo}, D.~J.
  {Champion}, D.~J. {Nice}, R.~{Ramachandran}, J.~W.~T. {Hessels},
  W.~{Vlemmings}, J.~{van Leeuwen}, S.~M. {Ransom}, N.~D.~R. {Bhat},
  Z.~{Arzoumanian}, M.~A. {McLaughlin}, V.~M. {Kaspi}, L.~{Kasian}, J.~S.
  {Deneva}, B.~{Reid}, S.~{Chatterjee}, J.~L. {Han}, D.~C. {Backer}, I.~H.
  {Stairs}, A.~A. {Deshpande}, and C.-A. {Faucher-Gigu{\`e}re}.
\newblock {Arecibo Pulsar Survey Using ALFA. I. Survey Strategy and First
  Discoveries}.
\newblock {\em \apj}, 637:446--455, January 2006.

\bibitem{deneva}
J.~S. {Deneva}, K.~{Stovall}, M.~A. {McLaughlin}, S.~D. {Bates}, P.~C.~C.
  {Freire}, J.~G. {Martinez}, F.~{Jenet}, and M.~{Bagchi}.
\newblock {Goals, Strategies and First Discoveries of AO327, the Arecibo
  All-sky 327 MHz Drift Pulsar Survey}.
\newblock {\em \apj}, 775(1):51, Sep 2013.

\bibitem{stovall}
K.~{Stovall}, R.~S. {Lynch}, S.~M. {Ransom}, A.~M. {Archibald}, S.~{Banaszak},
  C.~M. {Biwer}, J.~{Boyles}, L.~P. {Dartez}, D.~{Day}, A.~J. {Ford},
  J.~{Flanigan}, A.~{Garcia}, J.~W.~T. {Hessels}, J.~{Hinojosa}, F.~A. {Jenet},
  D.~L. {Kaplan}, C.~{Karako-Argaman}, V.~M. {Kaspi}, V.~I. {Kondratiev},
  S.~{Leake}, D.~R. {Lorimer}, G.~{Lunsford}, J.~G. {Martinez}, A.~{Mata},
  M.~A. {McLaughlin}, M.~S.~E. {Roberts}, M.~D. {Rohr}, X.~{Siemens}, I.~H.
  {Stairs}, J.~{van Leeuwen}, A.~N. {Walker}, and B.~L. {Wells}.
\newblock {The Green Bank Northern Celestial Cap Pulsar Survey. I. Survey
  Description, Data Analysis, and Initial Results}.
\newblock {\em \apj}, 791:67, August 2014.

\bibitem{siemens}
Xavier {Siemens}, Justin {Ellis}, Fredrick {Jenet}, and Joseph~D. {Romano}.
\newblock {The stochastic background: scaling laws and time to detection for
  pulsar timing arrays}.
\newblock {\em Classical and Quantum Gravity}, 30(22):224015, Nov 2013.

\bibitem{2018AmJPh..86..755L}
M.~T. {Lam}, J.~D. {Romano}, J.~S. {Key}, M.~{Normandin}, and J.~S. {Hazboun}.
\newblock {An acoustical analogue of a galactic-scale gravitational-wave
  detector}.
\newblock {\em American Journal of Physics}, 86:755--764, October 2018.

\bibitem{conversus}
J.~D. Romano.
\newblock Detector de {OG} de escala gal\'actica.
\newblock {\em Conversus}, 116:8--9, 2015.

\bibitem{Aloisi19}
R.~J. {Aloisi}, A.~{Cruz}, L.~{Daniels}, N.~{Meyers}, R.~{Roekle},
  A.~{Schuett}, J.~K. {Swiggum}, M.~E. {DeCesar}, D.~L. {Kaplan}, and R.~S.
  {Lynch}.
\newblock {The Green Bank North Celestial Cap Pulsar Survey. IV. Four New
  Timing Solutions}.
\newblock {\em \apj}, 875(1):19, Apr 2019.

\end{thebibliography}

\end{document}